# Personalized White Matter Bundle Segmentation for Early Childhood

Elyssa M. McMaster*[a], Michael E. Kim[b], Nancy R. Newlin[b], Gaurav Rudravaram[a], Adam M. Saunders [a], Aravind R. Krishnan[a], Jongyeon Yoon[b], Ji S. Kim[c], Bryce L. Geeraert[d], Meaghan V. Perdue[e], Catherine Lebel[d], Daniel Moyer[b], Kurt G. Schilling[f], Laurie E. Cutting [a,f,g,h], and Bennett A. Landman[a,b,f,g,h]

[a]Dept. of Electrical and Computer Engineering, Vanderbilt University, Nashville, TN, USA; [b]Dept. of Computer Science, Vanderbilt University, Nashville, TN, USA; [c]Creighton University School of Medicine, Omaha, NE, USA; [d]Dept. of Radiology, Cumming School of Medicine, Calgary, Canada; [e]Dept. of Psychiatry, UMass Chan Medical School, Worcester, MA, USA; [f]Dept. of Radiology and Radiological Sciences, Vanderbilt University Medical Center, Nashville, TN, USA; [g]Vanderbilt Kennedy Center, Vanderbilt University, Nashville, TN, USA; [h]Peabody College of Education, Vanderbilt University, Nashville, TN, USA

## ABSTRACT

White matter segmentation methods from diffusion magnetic resonance imaging range from streamline clustering-based approaches to bundle mask delineation, but none have proposed a pediatric-specific approach. White matter tract segmentation tools optimized for adults have shown inconsistent performance in pediatric populations. We explore optimization of manually corrected masks integrated into a direct-set, deep learning framework for generalizable pediatric white matter tract segmentation. We hypothesize that a deep learning model with a similar approach to TractSeg will improve similarity between an algorithm-generated mask and an expert-labeled ground truth. Given a cohort of 56 manually labelled white matter bundles, we take inspiration from TractSeg's 2D UNet architecture, and we modify (a) inputs to match bundle definitions as determined by pediatric experts; (b) evaluation to use k-fold cross validation instead of a train/validation/test split; (c) the loss function to masked Dice loss. We evaluate Dice score, volume overlap, and volume overreach of 16 major regions of interest compared to the expert-labeled dataset. To test whether our approach offers statistically significant improvements over TractSeg, we compare Dice voxels, volume overlap, and adjacency voxels with a Wilcoxon signed-rank test followed by false discovery rate (FDR) correction. We find statistical significance ($p < 0.05$) across all bundles for all metrics with one exception in volume overlap. After we run TractSeg and our model, we combine their output masks into a 60-label atlas to evaluate if TractSeg and our model combined can generate a robust, individualized atlas, and observe smoothed, continuous masks in cases that TractSeg did not produce an anatomically plausible output. With the improvement of white matter pathway segmentation masks, we can further understand neurodevelopment on a population-level scale, and we can produce reliable estimates of individualized anatomy in pediatric white matter diseases and disorders.

**Keyword**s: Segmentation, deep learning, diffusion MRI, pediatrics
*Elyssa.m.mcmaster@vanderbilt.edu

## 1. INTRODUCTION

Diffusion-weighted MRI (DW-MRI) is a non-invasive imaging modality to capture structural brain signal in vivo. Quantitative models characterize brain macrostructure and microstructure, such as diffusion tensor imaging (DTI) metrics[1,2], NODDI[3], and fiber orientation distribution functions (FODFs), which can be used to reconstruct virtual representations (streamlines) of white matter pathways in a process called tractography[1,4–7]. Appropriate delineation of white matter bundles paired with quantitative white matter models or tractography informs biomarkers of neurodevelopment on a population-level scale and produce estimates of individualized white matter alterations throughout the lifespan, including those with onset in early childhood, such as autism spectrum disorder (ASD)[8,9], attention deficit hyperactivity disorder (ADHD)[10], reading disabilities[11], and neurofibromatosis type 1 (NF1)[12,13].

Quantitative diffusion MRI metrics have paved the way for significant understanding of white matter development in specific bundles. Dimond et al. found developmental increases in fractional anisotropy (FA) and decreases in mean diffusivity (MD) across major white matter tracts[14]. Reynolds et al. found that the occipital and limbic connections develop significantly in early childhood and observed sex-related differences in development rate with tensor metrics[15]. Yu et al. investigated tensor metrics as a biomarker for ASD across different bundles[16], and Kumpulainen et al. investigated asymmetry and age-related white matter changes between infants and early childhood[13,17]. Meissner et al. characterized myelination and maturation in pediatrics and adults with white matter tractography and myelin water imaging[18]. Despite their compelling contributions to the understanding of the human brain, these studies have been limited to time-consuming manual segmentations[14,15,19] or white matter templates built for adult populations[16–18].

White matter tractography has great potential to accurately characterize individualized anatomy[20]. However, limitations of availability of expert-labelled bundle masks, data availability, and data quality from pediatric populations limits this approach in an automatic, generalizable context. In many cases, tools are built with a general population in mind while excluding out-of-distribution data and evaluation (Table 1).

Table 1: Literature review with popular existing white matter bundle segmentation method. Existing methods generate reasonable results for subjects above the age ranges of these studies, but no approach has been optimized for early childhood.

| White matter segmentation algorithm | Approach | Publication | Age group |
|---|---|---|---|
| RecoBundles | Streamline-based registration and clustering | Garyfallidis et al., 2017 | HCP-YA (Age 20-35) |
| TRACULA | Atlas-based probabilistic tracking with anatomical priors | Yendiki et al., 2011 | MGH in-house cohort (Age 27-52) |
| DeepWMA | CNN fiber classification | Zhang et al., 2020 | HCP-YA (Age 20-35) |
| TractSeg | U-Net to segment tractography-derived bundles | Wasserthal et al., 2018 | HCP-YA (20-35) |

White matter segmentation methods range from atlases[21] to streamline clustering-based methods[22] and semi-automatic tract extraction-turned-mask generation[23], but few have proposed a multi-subject, pediatric-specific approach. One approach proposed by Jordan et al. in 2021 takes inspiration from RecoBundles[22] to segment streamlines with an atlas, but it relies on an atlas generated from young adult anatomy; this approach may limit the possibility to reach upper-bound anatomical accuracy as otherwise determined by the literature[24]. Dimond et al. and Reynolds et al. employed semi-automatic white matter tract segmentation, but this process involves extensive manual labor and hours of quality control[14,15]. Yu et al. and Kumpulainen et al. employed the ICBM DTI-81 Atlas[25] based on 81 adults age 18 to 59[16,17]. Meissner et al. employed TRACULA[21], an approach based on subjects age 27-52[18]. Though these works make significant contributions to white matter development literature, it is an open question if they can benefit from a pediatric-specific approach. To date, leading white matter segmentation methods have been optimized on young adults and evaluated in aging and diseased populations (e.g., Alzheimer's disease cohorts or cases with tumors)[26]. TractSeg, a 2D UNet framework based on semi-automatic tractography-derived white matter segmentations, outperforms other white matter bundle segmentation methods and offers more bundles compared to its predecessors[23,27–29]. Recently, TractSeg has been shown to exhibit inconsistent performance in pediatric populations[30]. Considering the substantial changes undergone by the brain during development, we ask if the first white matter pathway segmentation algorithm tailored for a pediatric population can improve agreement with expert-labelled white matter masks (Figure 1)[31].

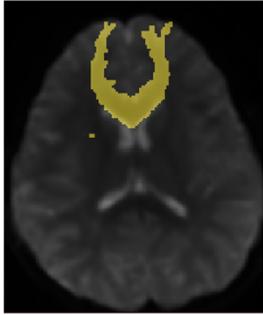

Figure 1: A pediatric brain with expert-labeled masks (top) and adult TractSeg-derived masks of the corpus callosum – genu (CC_Genu) and the left inferior occipito-frontal fascicle (L_IFO). Despite the well-established literature regarding the differences between pediatric and adult brains, no white matter segmentation tool to date caters specifically to pediatrics. Existing white matter bundle segmentation tools such as TractSeg not only focus on young adult anatomy, but high-quality acquisitions with isotropic voxels as well. In addition to exhibiting high variability in pediatrics, TractSeg also fails to produce segmentations in some anisotropic data, especially in smaller bundles, as noted in their GitHub README (https://github.com/MIC-DKFZ/TractSeg).

We propose a white matter tract segmentation algorithm designed and trained for pediatric subjects ages 2 through 8 based on the recent development of an expert-labelled ground truth dataset[32] from experts at the University of Calgary to overcome the current lack of a pediatric-specific white matter segmentation algorithm. We hypothesize that a deep learning model inspired by TractSeg and modified to handle the unique challenges presented by a small pediatric cohort will improve similarity between an algorithm-generated mask and an expert-labelled ground truth mask to approach upper-bound accuracy for anatomical representations; specifically, with literature-informed preprocessing steps and a lightweight deep learning approach, Dice voxels, volume overlap, volume overreach, bundle surface area, volume, average streamline length, and curl between expert-labelled and generated masks will improve compared to a leading approach. Though TractSeg does not generalize well to out-of-distribution pediatric data, we hypothesize that an additional model trained on the expert-labeled masks and TractSeg-derived masks will generalize across our population to produce continuous and qualitatively plausible segmentations.

## 2. METHODOLOGY

We obtain diffusion MRI data from 56 subjects in deidentified form from the Calgary preschool dataset[32] interpolated to 0.78 x 0.78 x 2.22 resolution with shells 0 and 750 mm/s$^2$. Subjects range in age from 2 years to 8 years at the time of the scan. We also obtain 16 regions of interest selected and segmented by experts in pediatric neuroimaging at the University of Calgary, including the body (CC_Body), splenium (CC_Splenium), and genu (CC_Genu) of the corpus callosum, the left and right cingulum (L_Cingulum and R_Cingulum), left and right superior longitudinal fasciculus (L_SLF and R_SLF), the left and right inferior occipito-frontal fascicle (L_IFO, R_IFO), the right and left pyramidal (R_Pyramidal and L_Pyramidal), the Fornix, the left and right inferior longitudinal fascicilus (L_ILF, R_ILF), and the left and right uncinate (L_Uncinate, R_Uncinate). Since pediatric experts labelled input masks, the bundles differ from current adult white matter segmentation tools due to differences in data quality, anatomy, and motivation. Experts at the University of Calgary generated masks from ExploreDTI[33] in the form of probability distributions for white matter pathway segmentations in each subject's anatomy.

Diffusion images underwent denoising, distortion correction, and quality assurance through the PreQual pipeline[34]. After preprocessing, we resampled diffusion images with cubic interpolation and masks with nearest neighbor interpolation to a common 1 mm isotropic in line with best practice for downstream processing as done in previous work[35,36]. We compute FODF peaks on the 1 mm isotropic images as our model's input, normalize peaks, and set NaN values to zero. We aggregate bundle masks into a single multi-volume image to allow our model to train on all bundles at once. We binarize white matter masks with a threshold of 0.5. We do not input mask slices with zero positive voxels for all bundles.

We implement a 2D U-Net inspired by TractSeg's approach[37]. Each input image has dimensions 200 x 200 x 119 x 9. We input a 2D image from each slice of the subject's axial plane (Figure 2). The target is a 2D multi-channel probability mask for each bundle based on the axial view of the image. The network consists of 5 encoder blocks, a bottleneck block, and 4 decoder blocks, with skip connections between encoder and decoder. Each block includes two convolutional layers followed by batch normalization and ReLU activation. Sigmoid activation is applied channel-wise in the final layer to produce voxel-wise probabilities for multi-label segmentation.

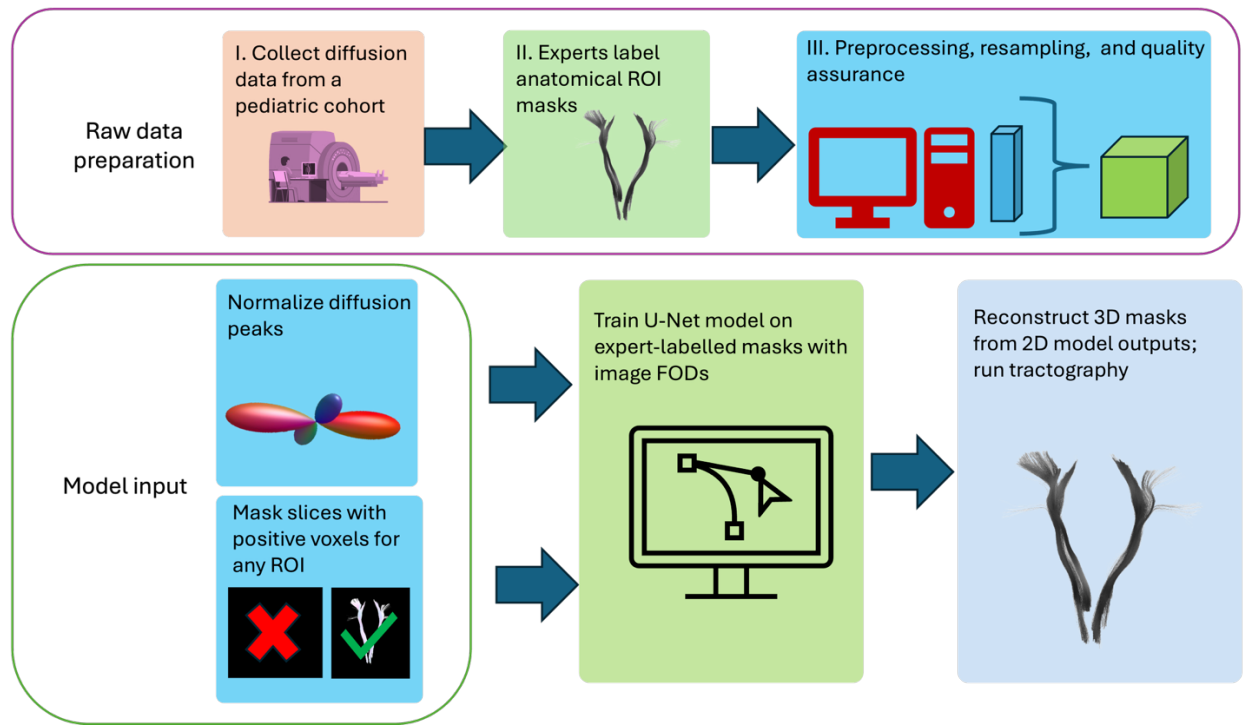

Figure 2: Following diffusion data collection from subjects age 2-8, the Calgary team extracted expert-labelled bundle segmentations. We compute FOD peaks as input to the model in an approach like TractSeg's, and load our target labels in a 4D, 16-volume file for streamlined training. We run inference on our validation sets for every fold and reconstruct 3D masks from the model's 2D outputs for every bundle.

The network implements a modified Dice loss function inspired by Feng et al. for sparse segmentation[38]. Specifically, we apply a spatial mask to prevent unlabeled or invalid voxels, such as background voxels or missing annotations, from contributing to the loss calculation. We compute average loss across all bundles for 250 epochs and stop early if the loss has not improved for 25 consecutive epochs. We set learning rate to 1e-3 with an Adamax optimizer. A sigmoid activation function is applied to the output logits, and probabilities are thresholded to 0.5 to produce binary masks. We perform 5-fold cross validation at the subject level. For each fold, we use the best model weights saved during training (selected on highest average validation Dice) to run inference on only that fold's validation subjects to avoid data leakage. After completing all folds, predictions were aggregated across all subjects to compute overall cross-validated performance. We reconstruct 3D volumes from the model's 2D outputs.

We compare the 3D white matter masks to the expert-annotated masks with Dice voxels, volume overlap, and volume overreach. To compare to a current leading approach, we run TractSeg out-of-the-box to generate equivalent bundle masks. However, TractSeg's high-quality adult training data allowed for a finer parcellation of bundles, and therefore, several TractSeg bundles must be merged for our equivalent segmentations. In high-resolution adult data like HCP-YA (TractSeg's training data), a fine parcellation of multiple segmentations of the same bundle may be ideal, but in lower quality acquisitions of pediatric brains, the same bundles may not be attainable. For this reason, we combine multiple TractSeg outputs into a single mask to cover the same bundle in pediatrics as our dataset – the CC_3 (Rostral body), CC_4 (anterior midbody), and CC_5 (posterior midbody) combined make up our CC_Body equivalent; the SLF_I, SLF_II, and SLF_III for left and right are combined to be equivalent to our R_SLF and L_SLF. We perform the same Dice voxels, volume overlap, and volume overreach computations between the expert-labeled masks and the TractSeg-generated masks. We perform a Wilcoxon Signed-Rank test between the TractSeg distribution and the proposed model's distribution with false discovery rate (FDR) correction to establish a statistical difference in similarity from the expert-labeled dataset. We perform probabilistic iFOD2 tracking across the expert-labeled bundle masks and the proposed model's outputs. Bundles with the same requirements – minimum streamline length 40 mm, maximum streamline length 200 mm, cutoff of 0.1, and 2000 generated streamlines[14,30].

Though we do not have expert-labelled masks for every TractSeg bundles, we leverage our model's architecture to generate anatomically informed, continuous masks from those bundles. With the same architecture, we change from 16 output channels to 60 to append the following TractSeg bundles absent from our expert-labelled dataset in Table 2. In the absence of a ground truth, we evaluate these results qualitatively for anatomical feasibility.

Table 2: Bundles appended to our inputs after TractSeg segmentation on the pediatric dataset.

| Name | Abbreviation |
| --- | --- |
| Arcuate fascicle | AF_left, AF_right |
| Anterial thalamic radiation | ATR_left, ATR_right |
| Commissure anterior | CA |
| Corticospinal tract | CST_left, CST, right |
| Fronto-pntine tract | FPT_left, FPT_right |
| Inferior cerebellar peduncle | ICP_left, ICP_right |
| Middle cerebellar peduncle | MCP |
| Middle longitudinal fascicle | MLF_left, MLF_right |
| Optical radiation | OR_left, OR_right |
| Parieto-occipital pontine | POPT_left, POPT_right |
| Superior cerebellar peduncle | SCP_left, SCP_right |
| Superior thalamic radiation | STR_left, STR_right |
| Striato-fronto-orbital | ST_FO_left, ST_FO_right |
| Striato-prefrontal | ST_PREF_left, ST_PREF_right |
| Striato-precentral | ST_PREC_left, ST_PREC_right |
| Striato-parietal | ST_PAR_left, ST_PAR_right |
| Striato-occipital | ST_OCC_left, ST_OCC_right |
| Stratio-postcentral | ST_POSTC_left, ST_POSTC_right |

| Thalamo-premotor | T_PREM_left, T_PREM_right |
| Thalamo-precentral | T_PREC_left, T_PREC_right |
| Thalamo-postcentral | T_POSTC_left, T_POSTC_right |
| Thalamo-parietal | T_PAR_left, T_PAR_right |
| Thalamo-occipital | T_OCC_left, T_OCC_right |

## 3. RESULTS

Our analysis relies on comparison between existing expert-labelled masks with successfully generated masks from both TractSeg and the proposed model. Our dataset includes one subject with a missing L_ILF, one with missing L_IFO, two with missing L_Uncinate, one with missing Fornix, and one with missing R_Uncinate. We do not report results for the R_ILF due to missing bundles for 22 subjects. Within the shared bundles, TractSeg could not generate any part of the Fornix for 29 subjects. We exclude the L_Pyramidal and R_Pyramidal as TractSeg does not include these bundles in its parcellation.

Following our Wilcoxon signed-rank test, computed on a bundle-wise and metric-wise basis, we find statistical significance in every case for Dice voxels and adjacency voxels; for volume overlap, only the R_Cingulum did not exhibit a positive result. We present the Dice scores with statistical significance between the proposed model and the TractSeg outputs in Figure 3.

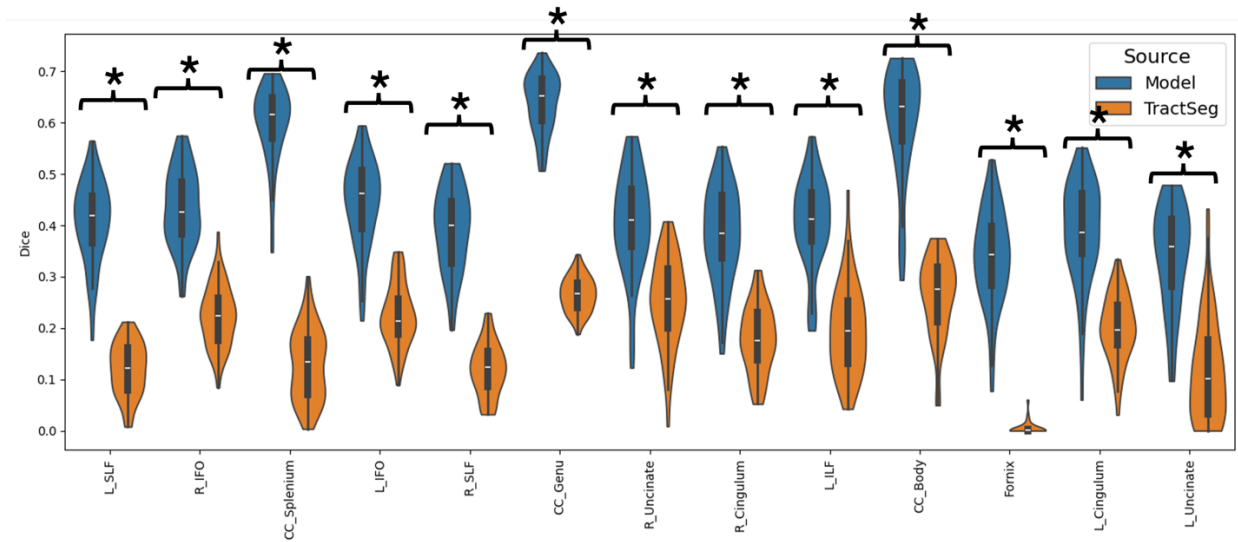

Figure 3: We present the Dice scores between the expert-labelled ground truth and the proposed model's outputs (* indicates statistical significance where $p < 0.05$ from a Wilcoxon Signed-Rank Test followed by FDR correction). Even in the worst cases, we observe substantial improvement compared to TractSeg's predicted segmentations for the pediatric population. Notably, TractSeg fails to segment the Fornix in many cases, especially in anisotropic-acquired data, due to its small size, but the proposed model shows significant progress toward a pediatric Fornix segmentation.

Qualitatively, we observe continuous, anatomically plausible masks from the proposed model's outputs, as opposed to fragmented outputs from the same bundles from TractSeg (Figure 4). Even in bundles with lower Dice agreement between the expert-labelled masks and the proposed model's outputs (the L_Uncinate, for example), we still observe smooth, anatomically feasible white matter segmentations that have potential to generate white matter tractography in that region. For our analysis, we focus on one large bundle that significantly outperforms TractSeg in Dice scores (the CC_Body), one medium-size bundle that outperforms TractSeg but still exhibits a low Dice score (R_SLF), and one small bundle with Dice scores that approach TractSeg's outputs (L_Uncinate).

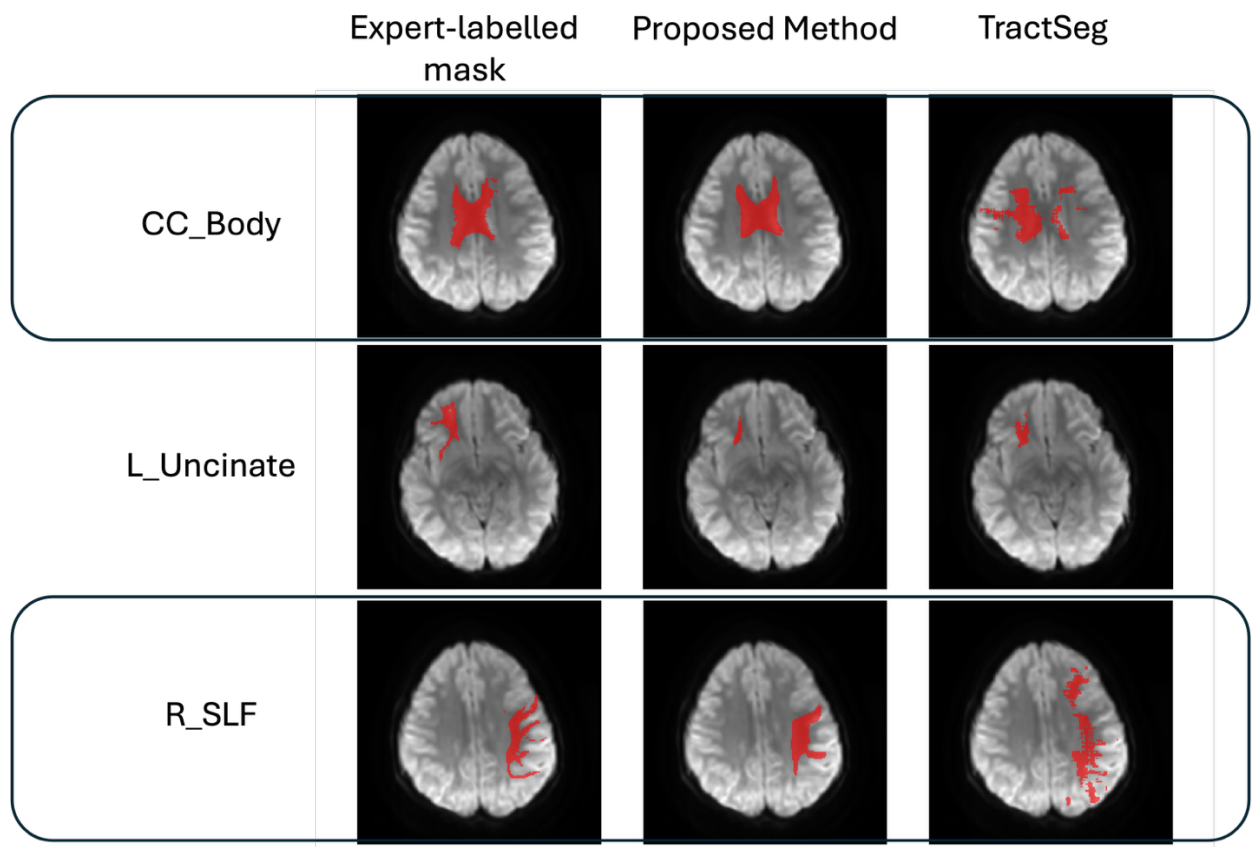

Figure 4: We show the qualitative results for white matter bundle segmentation from the proposed model and TractSeg as compared to the expert-labelled ground truth. Though the proposed model's outputs appear overly smooth compared to the expert-labelled mask, the smoothness allows for a continuous mask, which is required to run *tckgen* across a bundle.

We run iFOD2 tractography with *tckgen* across our proposed model's outputs and the expert-labeled masks. We display the absolute value of effect size on a heatmap to compare bundles in Figure 5. As defined in the literature, effect size of 0.2 = small effect size, 0.5 = medium effect size, and 0.8 = large effect size[39], the CC_Body and L_Uncinate display only small and medium effect sizes, but the R_SLF's surface area and volume show large effect sizes compared to the expert-labeled masks.

We investigate the qualitative results of our 60-bundle inputs of expert-labelled masks combined with TractSeg outputs in our model with the AF_left, CST_right, and OR_left, as these pathways represent different brain region connections and appear frequently in anatomical and neuroimage processing literature[36,40–43] . Overall, we observe smoother masks that appear more anatomically plausible than the TractSeg outputs alone in Figure 6.

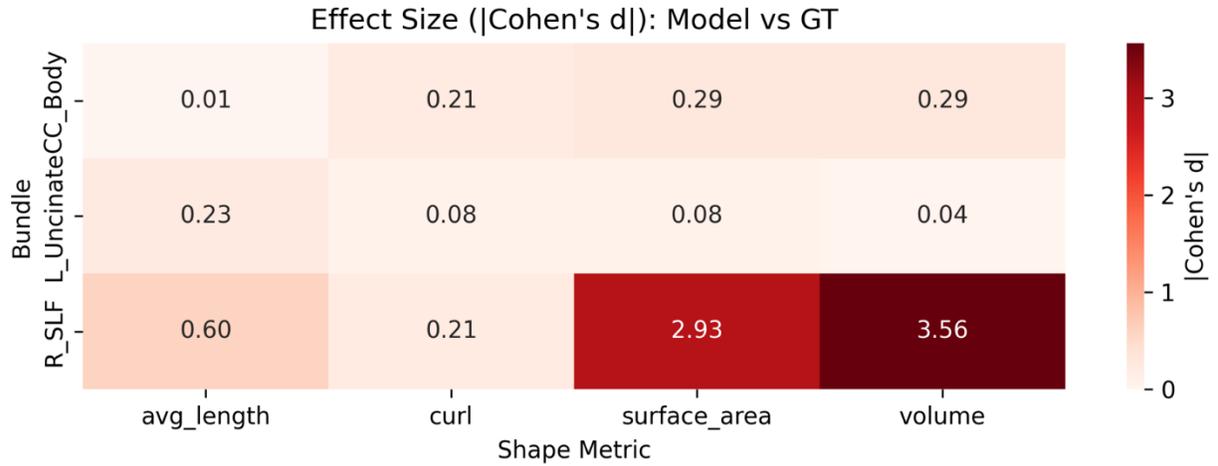

Figure 5: Cohen's d effect size (absolute value) between our three bundles in white matter tractography metrics for our model's outputs and the expert-labelled ground truth mask (GT). Despite similar Dice scores between distributions, the L_Uncinate shows small effect size between tractography measures while the R_SLF's effect sizes indicate that the means of the two distributions differ more than their standard deviations despite similar average streamline length and curl metrics.

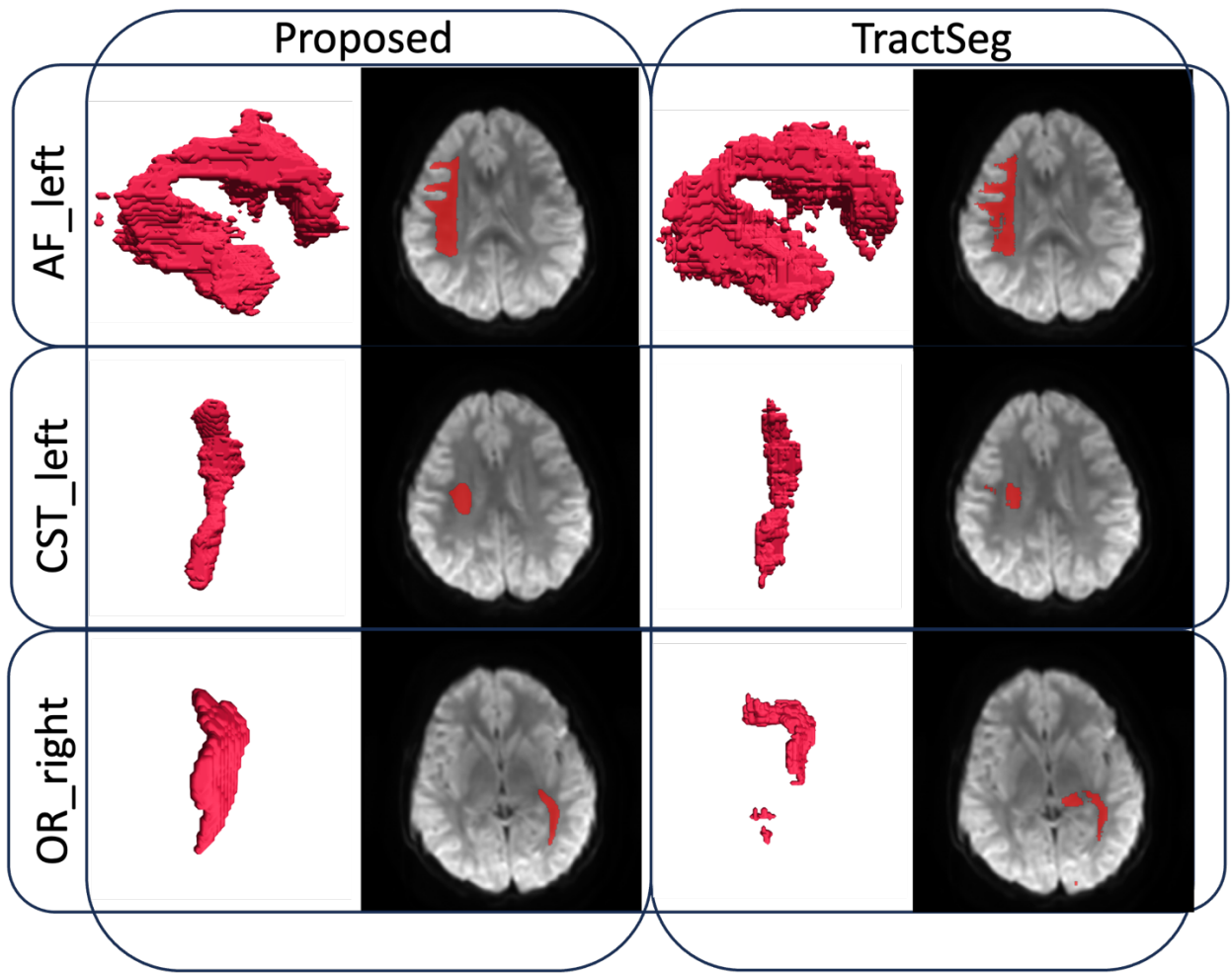

Figure 6: We display the proposed model's outputs for the CST_left, the OR_right, and the AF_left given only the subject's TractSeg outputs to train on. We observe greater continuity in the masks despite fragmented training data.

## 4. DISCUSSION

Overall, we observe a statistically significant improvement across mask similarity measures with the proposed model's outputs compared to TractSeg. Though some bundles exhibit suboptimal Dice scores (<0.5 Dice) compared to the expert-labelled masks, these segmentations still outperform TractSeg. Notably, even in the case of poor Dice scores, the output masks show higher qualitative continuity; mask continuity is essential for successful bundle tractography. In our study of three bundles, only the SLF showed large effect size in any white matter tractography metric. An additional consideration for this study includes spatial sampling – though we resample the diffusion images and binary masks to the recommended 1 mm isotropic resolution, the original bundle masks were generated in the image's original anisotropic resolution before resampling. Perhaps with a higher resolution original mask, we could achieve a finer parcellation and higher agreement between mask metrics for the expert-labelled ground truth and the proposed model. Additionally, a pre-training check that an input slice contains positive voxels on a bundle-wise basis instead of positive voxels for any bundle may be more appropriate preprocessing. The loss function could also be updated to handle data on a per-bundle basis instead of per-batch.

We also developed a 60-bundle atlas trained on 16 expert-labelled bundle segmentations and 44 TractSeg-derived masks. Though we show that TractSeg exhibits low Dice performance on pediatric subjects as compared to the proposed model, it served here as a useful anatomical prior for this demographic; a first step as an input to a model that follows by smoothing the bundle as appropriate for the population. A major advantage of TractSeg over its predecessors is its high number of masks, and integration of their bundles in a pediatric cohort remains an open avenue for future research.

Through this work, we have shown that pediatric white matter segmentation requires an age-appropriate atlas. With the successful generation of binary white matter masks for pediatric subjects, we can extract meaningful statistics within these bundles for DTI metrics, NODDI metrics, and tractography.

## ACKNOWLEDGEMENTS

The Vanderbilt Institute for Clinical and Translational Research (VICTR) is funded by the National Center for Advancing Translational Sciences (NCATS) Clinical Translational Science Award (CTSA) Program, Award Number 5UL1TR002243-03. This work was conducted in part using the resources of the Advanced Computing Center for Research and Education (ACCRE) at Vanderbilt University, Nashville, TN, as well as NIH 5U01DA055347-03, ADSP U24AG074855, DoD grant HT94252410563 and NIH 1R01EB017230. This work was supported by the Alzheimer's Disease Sequencing Project Phenotype Harmonization Consortium (ADSP-PHC) that is funded by NIA (U24 AG074855, U01 AG068057 and R01 AG059716). The content is solely the responsibility of the authors and does not necessarily represent the official views of the NIH.

We used generative artificial intelligence (AI) to create code segments based on task descriptions, as well as to debug, edit, and autocomplete code. Additionally, generative AI technologies have been employed to assist in structuring sentences and performing grammatical checks. The conceptualization, ideation, and all prompts provided to the AI originated entirely from the authors' creative and intellectual efforts. We take accountability for the review of all content generated by AI in this work.

This work has not been submitted for publication or presentation elsewhere.